\def\be{\begin{equation}}
\def\ee{\end{equation}}
\def\ba{\begin{array}}
\def\ea{\end{array}}

\documentclass[aps,amsmath,amssymb,amsfonts,showpacs]{revtex4}
\usepackage{graphicx}
\usepackage{epstopdf}
\def\qed{\leavevmode\unskip\penalty9999 \hbox{}\nobreak\hfill
     \quad\hbox{\leavevmode  \hbox to.77778em{%
               \hfil\vrule   \vbox to.675em%
               {\hrule width.6em\vfil\hrule}\vrule\hfil}}
     \par\vskip3pt}

\newtheorem{theorem}{Theorem}
\newtheorem{corollary}{Corollary}
\newtheorem{lemma}{Lemma}

\begin{document}
\title{ Quantifying quantum coherence based on the generalized  $\alpha-z-$relative R$\acute{e}$nyi entropy}
\author{Xue-Na Zhu$^{1}$}
\author{Zhi-Xiang Jin$^{2}$}
\author{Shao-Ming Fei$^{3,4}$}

\affiliation{$^1$School of Mathematics and Statistics Science, Ludong University, Yantai 264025, China\\
$^2$School of Physics, University of Chinese Academy of Sciences, Yuquan Road 19A, Beijing 100049, China\\
$^3$School of Mathematical Sciences, Capital Normal University,
Beijing 100048, China\\
$^4$Max-Planck-Institute for Mathematics in the Sciences, 04103
Leipzig, Germany}

\begin{abstract}
We present a family of coherence quantifiers based on the generalized $\alpha-z-$relative R$\acute{e}$nyi entropy.
These quantifiers satisfy all the standard criteria for well-defined measures of coherence, and include some existing coherence measures as
special cases.
\end{abstract}

\maketitle

\section{Introduction}
Coherence, being at the heart of interference
phenomena, plays a central role in quantum physics as it
enables applications that are impossible within classical
mechanics or ray optics.
Coherence is also a vital physical resource with various
applications in biology \cite{c1,c2,c3}, thermodynamical systems \cite{c4,c5}, transport theory \cite{c6,c7} and nanoscale physics \cite{c8}.
Recent developments in our understanding of quantum coherence \cite{K,A,X,C,X2,S} and nonclassical
correlation have come from the burgeoning field of quantum information science.
One important pillar of the field is the study on quantification of coherence.

In Ref. \cite{PRL113.140401} the authors established a rigorous framework (BCP framework) for quantifying coherence. The BCP framework consists of the following postulates
that any quantifier of coherence $C$ should fulfill:

$(C_1)$ Faithfulness:
$
C(\rho)\geq0,
$
with equality if and only if $\rho$ is incoherent.

$(C_2)$ Monotonicity: $C$ does not increase under the action
of an incoherent operation, i.e.,
\begin{equation}\nonumber
C[\Phi_{\mathcal{I}}(\rho)]\leq C(\rho),
\end{equation}
for any
incoherent operation $\Phi_{\mathcal{I}}$.

$(C_3)$ Convexity: $C$ is a convex function of the state, i.e.,
\begin{equation}\nonumber
\sum_np_n C(\rho_n)\geq C(\sum_np_n\rho_n),
\end{equation}
where $p_n\geq0, \sum_np_n=1$.

$(C_4)$ Strong monotonicity: $C$ does not increase on average
under selective incoherent operations, i.e,
\begin{equation}\nonumber
C(\rho)\geq\sum_np_n C(\varrho_n),
\end{equation}
with probabilities $p_n=tr(\mathcal{K}_n\rho\mathcal{K}^{\dag}_n)$,
post measurement states
$\varrho_n=\frac{\mathcal{K}_n\rho\mathcal{K}^{\dag}_n}{p_n},$
and incoherent operators ${\mathcal{K}}_n$.

The authors of Ref. \cite{PRA94060302} provided a simple and interesting
condition to replace (C3) and (C4) with the additivity of
coherence for block-diagonal states,
\begin{equation}\label{+}
C(p\rho_1\oplus(1-p)\rho_2)=pC(\rho_1)+(1-p)C(\rho_2),
\end{equation}
for any $p\in[0,1]$, $\rho_i\in\varepsilon(\mathcal{H}_i)$, $i=1, 2$,
and $p\rho_1\oplus(1-p)\rho_2\in\varepsilon(\mathcal{H}_1\oplus \mathcal{H}_2),$
where $\varepsilon(\mathcal{H})$ denotes the set of density
matrices on the Hilbert space $\mathcal{H}$.

For a given $d$-dimensional Hilbert space $\mathcal{H}$, let us fix an orthonormal basis $\{|i\rangle\}_{i=1}^{d}$. We call all density matrices that are
diagonal in this basis incoherent and label this
set of quantum states by $\mathcal{I}\subset\mathcal{H}$. All density
operators $\delta\in \mathcal{I}$ are of the form:
\begin{equation}\nonumber
\delta=\sum_ip_i|i\rangle\langle i|,
\end{equation}
where $p_i\geq0$ and $\sum_ip_i=1$.
Otherwise the states are coherent.
Let
$\Lambda$ be a completely positive trace preserving (CPTP) map:
\begin{equation}\nonumber
\Lambda(\rho)=\sum_i\mathcal{K}_n\rho \mathcal{K}^{\dag}_n,
\end{equation}
where $\{\mathcal{K}_n\}$ is a set of Kraus operators satisfying
$\sum_n\mathcal{K}^{\dag}_n\mathcal{K}_n=I_d$, with $I_d$ the identity operator. If $\mathcal{K}^{\dag}_n\mathcal{I}\textit{K}_n\in\mathcal{I}$ for all $n$, we call $\{\mathcal{K}_n\}$ a set of incoherent Kraus
operators, and the corresponding operation $\Lambda$ an incoherent operational one.

\section{the function $f_{\alpha,z}(\rho,\sigma)$}
Quantifying coherence is a key task in both quantum mechanical theory and practical applications.
In Ref. \cite{98,az} the following function has been presented,
\begin{equation}\label{f}
f_{\alpha,z}(\rho,\sigma)
=Tr(\sigma^{\frac{1-\alpha}{2z}}
\rho^{\frac{\alpha}{z}}\sigma^{\frac{1-\alpha}{2z}})^{z},
\end{equation}
for arbitrary two density matrices $\rho$ and $\sigma$.
Here, $\alpha,\,z\in R$. To study the limit when $\alpha\rightarrow 1$ and $z\rightarrow 0$, the authors in Ref. \cite{az} parameterized $z$ in terms of $\alpha$ as $z=r(\alpha-1)$, where $r$ is a non-zero
finite real number, and considered the limit when $\alpha\rightarrow 1$:
$\lim_{\alpha\rightarrow1}f_{\alpha,r(\alpha-1)}(\rho,\sigma)=\rho$. For fixed $\alpha\not=1$, $z\rightarrow 0$ is exactly related to the anti Lie-Trotter problem \cite{kmr}.

For a finite dimensional Hilbert space $\mathcal{H}$,
the set of linear operators is denoted by $\mathcal{L(H)}$.
The adjoint of $X\in \mathcal{L(H)}$ is denoted by $X^{\dag}$.
For $X\in \mathcal{L(H)}$ and real $p\not=0$,  $||X||_{p}$ is defined by [20],
\begin{equation}\nonumber
||X||_{p}
=(tr|X|^{p})^{\frac{1}{p}},
\end{equation}
where $|X|=\sqrt{X^{\dag}X}$. Here,
for a self-adjoint operator $X$, $X^{-1}$ means the inverse restricted to $supp(X)$,
so $X^{-1}X=XX^{-1}$ equals to the orthogonal projection on $supp(X)$.

The
$H\ddot{o}lder^{\prime}s$ inequality belongs to a richer family of inequalities. For every $p_1,..., p_k, r>0$ with
$\frac{1}{r}=\frac{1}{p_1}+...+\frac{1}{p_k}$ one has \cite{tr}:
\begin{equation}\label{x1}
||X_1...X_k||_{r}
\leq||X_1||_{p_1}...||X_k||_{p_k}.
\end{equation}
From this inequality and the fact that $||X^{-1}||_{-p}=||X||_{p}^{-1}$, the following reverse $H\ddot{o}lder^{\prime}s$ inequality
is derived.
Let $r>0$ and $p_1,..., p_k$ be such that
$\frac{1}{r}=\frac{1}{p_1}+...+\frac{1}{p_k}$
and that exactly one of $p_i^{\prime}$s is
positive and the rests are negative \cite{tr}:
\begin{equation}\label{x2}
||X_1...X_k||_{r}
\geq||X_1||_{p_1}...||X_k||_{p_k}.
\end{equation}
Moreover, equalities  holds in (\ref{x1}) and (\ref{x2})
if and only if $|X_i|^{p_i}, i=1,2,...,k$, are proportional.

\begin{lemma}For states $\rho$ and $\sigma$,

(1) If $0<\alpha<1$ and $z>0$, we have
\begin{equation}\nonumber\label{f11}
f_{\alpha,z}(\rho,\sigma)\leq1;
\end{equation}

(2) If $\alpha>1$ and $z>0$, we have
\begin{equation}\nonumber\label{f12}
f_{\alpha,z}(\rho,\sigma)\geq 1.
\end{equation}

(3) $f_{\alpha,z}(\rho,\sigma)=1$ if and only if $\rho=\sigma,$
for $\alpha\in(0,1)\cup(1,+\infty)$ and $z>0$.

\end{lemma}

{\sf [Proof]} Let $r=z$,
$p_1=\frac{2z}{1-\alpha}$, $p_2=\frac{z}{\alpha},$
$X_{1}=\sigma^{\frac{1}{p_1}}$, $X_{2}=\rho^{\frac{1}{p_2}}$.
When $\alpha\in(0,1)$ and $z>0$, we have
\begin{eqnarray}\label{de1}
f_{\alpha,z}(\rho,\sigma)&=&tr(X_1X_2X_1)^z\\\nonumber
&=&tr(|X_1X_2X_1|)^{r}\\\nonumber
&=&(||X_1X_2X_1||_{r})^{r}\\\nonumber
&\leq&(||X_1||_{p_1}||X_2||_{p_2}||X_1||_{p_1})^{r}\\\nonumber
&=&1,
\end{eqnarray}
where the second equality is due to $X_i^{\dag}=X_i$ for $i=1,2.$
From (\ref{x1}), we obtain the first inequality.

When $\alpha>1$ and $z>0$, we have
\begin{eqnarray}\label{de2}
f_{\alpha,z}(\rho,\sigma)
&=&(||X_1X_2X_1||_{r})^{r}\\\nonumber
&\geq&(||X_1||_{p_1}||X_2||_{p_2}||X_1||_{p_1})^{r}\\\nonumber
&=&1,
\end{eqnarray}
where
the first inequality is due to (\ref{x2}).

In  the above proof of inequalities (\ref{de1}) and (\ref{de2}), $||X_1X_2X_1||_{r}=||X_1||_{p_1}||X_2||_{p_2}||X_1||_{p_1}$
if and only if $|X_1|^{p_1}$ and $|X_2|^{p_2}$  are proportional, i.e, there is a  number $k$ which satisfies $\sigma=k\rho$. Since $tr(\rho)=tr(\sigma)=1$, then we obtain $k=1$. $\Box$

Let $\mathcal{P(H)}$ be the set of positive semidefinite operators on $\mathcal{H}$.
For non-normalized states $\rho$: $\forall\rho,\sigma\in \mathcal{P(H)}$
with $supp\,\rho\subseteq supp\,\sigma$, it has been defined in Ref. \cite{az},
\begin{equation}\label{DF}
D_{\alpha,z}(\rho||\sigma)
:=\frac{1}{\alpha-1}\log\frac{f_{\alpha,z}(\rho,\sigma)}{tr\rho}.
\end{equation}
For any states $\rho,\sigma$ such that $supp\,\rho\subseteq supp\,\sigma$, and for any CPTP map $\Lambda$:
$
D_{\alpha,z}(\Lambda(\rho)||\Lambda(\sigma))\leq
D_{\alpha,z}(\rho||\sigma)
$ holds in each of the following cases \cite{az}:

$\bullet$ $\alpha\in(0,1]$ and $z\geq\max\{\alpha,1-\alpha\}$;

$\bullet$
$\alpha\in[1,2]$ and $z=1$;

$\bullet$  $\alpha\in[1,2]$ and $z=\frac{\alpha}{2}$.

$\bullet$
$\alpha\geq1$ and $z=\alpha$.

For two states $\rho$ and $\sigma$, one has $f_{\alpha,z}(\rho,\sigma)=e^{(\alpha-1)D_{\alpha,z}(\rho||\sigma)}$.
Hence
$
f_{\alpha,z}(\rho,\sigma)
$ has the following  properties:

\begin{lemma}\label{Lf}
For any quantum states $\rho$ and $\sigma$, such that $supp\,\rho\subseteq supp\,\sigma$, and for any CPTP map $\Lambda$, we have

$\bullet$ If $\alpha\in(0,1]$ and $z\geq\max\{\alpha,1-\alpha\}$, then
\begin{equation*}
f_{\alpha,z}(\Lambda(\rho),\Lambda(\sigma))\geq
f_{\alpha,z}(\rho,\sigma);
\end{equation*}

$\bullet$ If
$\alpha\in[1,2]$ and $z\in\{1,\frac{\alpha}{2}\}$; or
$\alpha\geq1$ and $z=\alpha$, then
\begin{equation*}
f_{\alpha,z}(\Lambda(\rho),\Lambda(\sigma))\leq
f_{\alpha,z}(\rho,\sigma).
\end{equation*}
\end{lemma}

\section{Coherence quantification}
The coherence
$C(\rho)$ in Ref. \cite{PRA95.042337} can be expressed as
\begin{equation}\label{C_1}
C(\rho)=1-\left[\max_{\sigma\in\mathcal{I}}f_{\frac{1}{2},1}(\rho,\sigma)\right]^2.
\end{equation}
In Ref. \cite{Jinzhixiang1} a bona fide measure of quantum coherence $C(\rho)$ has been presented by utilizing the Hellinger distance: $D_H(\rho,\sigma)=Tr(\sqrt{\rho}-\sqrt{\sigma})^2$,
\begin{eqnarray}\label{C_2}
C(\rho)&=&\min_{\sigma\in \mathcal{I}}D_H(\rho,\sigma)\\\nonumber
&=&2\left[1-\max_{\sigma\in\mathcal{I}}f_{\frac{1}{2},1}(\rho,\sigma)\right],
\end{eqnarray}
which is the coherence $C_{\frac{1}{2}}(\varepsilon|\rho)$ of Theorem 3 in Ref. \cite{PRA93.032136}.

In Ref. \cite{PRA93.032136} the coherence has been quantified based on the Tsallis relative $\alpha$ entropy,
 \begin{equation}\label{Tq}
 D^{\prime}_{\alpha}(\rho||\sigma)
 =\frac{1}{\alpha-1}(f_{\alpha,1}(\rho,\sigma)-1).
 \end{equation}
But it was shown that it to violates the strong monotonicity, even though it can
unambiguously distinguish the coherent state from the incoherent ones with the monotonicity.
In Ref. \cite{R} a family of coherence quantifiers has been presented, which are closely related to the Tsallis relative $\alpha$ entropy:
\begin{equation}\label{C_3}
C^{\prime}_{\alpha}(\rho)
=\min_{\sigma\in \mathcal{I}}\frac{1}{\alpha-1}\left(f_{\alpha,1}^{\frac{1}{\alpha}}(\rho,\sigma)-1\right),
\end{equation}
where $\alpha\in(0,2].$

In the following we define a generalized $\alpha-z-$relative  $R\acute{e}nyi$ entropy:
\begin{equation}\label{D}
D_{\alpha,z}(\rho,\sigma)
=\frac{f_{\alpha,z}^{\frac{1}{\alpha}}(\rho,\sigma)-1}{\alpha-1}.
\end{equation}

It is worthwhile noting that several coherence measures like relative
entropy \cite{PRL113.140401}, geometric coherence \cite{2c}, the sandwiched $R\acute{e}nyi$ relative entropy \cite{JX} and max-relative
entropy \cite{K} are related to the generalized $\alpha-z-$relative  $R\acute{e}nyi$ entropy.

Based on the relation $f_{\alpha,z}(\rho,\sigma)$ and $D_{\alpha,z}(\rho,\sigma)$, and Lemma \ref{Lf},
we have
\begin{corollary}\label{Cor1}
For any quantum states $\rho$ and $\sigma$ for which $supp\,\rho\subseteq supp\,\sigma$, and for any CPTP map $\Lambda$:
$
D_{\alpha,z}(\Lambda(\rho),\Lambda(\sigma))\leq
D_{\alpha,z}(\rho,\sigma)
$ holds in each of the following case:

$\bullet$ $\alpha\in(0,1]$ and $z\geq\max\{\alpha,1-\alpha\}$;

$\bullet$
$\alpha\in[1,2]$ and $z=1$;

$\bullet$  $\alpha\in[1,2]$ and $z=\frac{\alpha}{2}$;

$\bullet$
$\alpha\geq1$ and $z=\alpha$.
\end{corollary}

With the above properties, based on  the generalized $\alpha-z-$relative  $R\acute{e}nyi$ entropy we define the quantity:
$C_{\alpha,z}(\rho)=\min_{\sigma\in \mathcal{I}}D_{\alpha,z}(\rho,\sigma)$.
The following statement takes place.

\begin{theorem}\label{th1}
 The quantum coherence $C_{\alpha,z}(\rho)$  of a state $\rho$ given by
 \begin{equation}\label{Ca}
C_{\alpha,z}(\rho)=\min_{\sigma\in \mathcal{I}}D_{\alpha,z}(\rho,\sigma)
\end{equation}
is a well-defined measure of coherence for the following case:

$\bullet$ $\alpha\in(0,1)$ and $z\geq\max\{\alpha,1-\alpha\}$;

$\bullet$
$\alpha\in(1,2]$ and $z=1$;

$\bullet$    $\alpha\in(1,2]$ and $z=\frac{\alpha}{2}$;

$\bullet$
$\alpha>1$ and $z=\alpha$.
\end{theorem}

{\sf [Proof]} Because of (\ref{f}), (\ref{D}) and (\ref{Ca}), we have
\begin{equation}\nonumber
C_{\alpha,z}(\rho)=\left\{\begin{array}{llrr}
\frac{1-\max_{\sigma\in \mathcal{I}}f^{\frac{1}{\alpha}}_{\alpha,z}(\rho,\sigma)}{1-\alpha},&0<\alpha<1,\\
\frac{\min_{\sigma\in \mathcal{I}}f^{\frac{1}{\alpha}}_{\alpha,z}(\rho,\sigma)-1}{\alpha-1},&\alpha>1.
\end{array}\right.
\end{equation}
From Lemma 1, we have $C_{\alpha,z}(\rho)\geq0$,
and $C_{\alpha,z}(\rho)=0$ if and only if $\rho=\sigma.$
Let $\sigma$ be the optimal incoherent state such that $C_{\alpha,z}(\rho)=D_{\alpha,z}(\rho,\sigma).$
Taking into account Corollary \ref{Cor1}, we have that
$C_{\alpha,z}(\rho)$ does not increase under any incoherent operations.

Next we prove that $C_{\alpha,z}(\rho)$ satisfies Eq. (\ref{+}).
Suppose $\rho$ is block-diagonal in the reference basis $\{|j\rangle\}_{j=1}^{d}$,
$\rho=p_1\rho_1\oplus p_2\rho_2$
with $p_1\geq0, p_2\geq0$, $p_1+p_2=1,$ $\rho_1$ and $\rho_2$ are density operators.
Let $\sigma=q_1\sigma_1\oplus q_2\sigma_2$
with $q_1\geq0, q_2\geq0$, $q_1+q_2=1,$ and $\sigma_1, \sigma_2$ are diagonal states similar to $\rho_1, \rho_2$, respectively.

Denote $\Delta$ either $\max$ or $\min$. Set $t_i=\Delta_{\sigma_i}tr(\sigma_i^{\frac{1-\alpha}{2z}}
\rho_i^{\frac{\alpha}{z}}\sigma_i^{\frac{1-\alpha}{2z}})^z$, $i=1, 2$. We have
\begin{eqnarray}\label{delta}
&&\Delta_{\sigma\in \mathcal{I}}
tr(\sigma^{\frac{1-\alpha}{2z}}
\rho^{\frac{\alpha}{z}}\sigma^{\frac{1-\alpha}{2z}})^{z}\\\nonumber
&=&\Delta_{q_1,q_2}(q_1^{1-\alpha}p_1^{\alpha}t_1+q_2^{1-\alpha}p_2^{\alpha}t_2).
\end{eqnarray}
Due to the H$\ddot{o}$lder inequality with $0<\alpha<1$, we have
\begin{eqnarray*}
q_1^{1-\alpha}p_1^{\alpha}t_1+q_2^{1-\alpha}p_2^{\alpha}t_2
\leq(\sum_{i=1,2}p_it_i^{\frac{1}{\alpha}})^{\alpha},
\end{eqnarray*}
where the equality holds if
and only $q_1=lp_1t^{\frac{1}{\alpha}}_1$
and $q_2=lp_2t^{\frac{1}{\alpha}}_2$ with $l=\left[p_1t^{\frac{1}{\alpha}}_1+
p_2t^{\frac{1}{\alpha}}_2\right]^{-1}$,
i.e, \begin{eqnarray}\label{max}
\max_{q_1,q_2}(q_1^{1-\alpha}p_1^
{\alpha}t_1+q_2^{1-\alpha}p_2^{\alpha}t_2)
=(\sum_{i=1,2}p_it_i^{\frac{1}{\alpha}})^{\alpha}.
\end{eqnarray}

Similarly, for the inequality with $\alpha>1$, we have
\begin{eqnarray*}
q_1^{1-\alpha}p_1^{\alpha}t_1+q_2^{1-\alpha}p_2^{\alpha}t_2
\geq(\sum_{i=1,2}p_it_i^{\frac{1}{\alpha}})^{\alpha}.
\end{eqnarray*}
When $q_1=lp_1t^{\frac{1}{\alpha}}_1$
and $q_2=lp_2t^{\frac{1}{\alpha}}_2$, we obtain
\begin{eqnarray}\label{min}
\min_{q_1,q_2}(q_1^{1-\alpha}p_1^
{\alpha}t_1+q_2^{1-\alpha}p_2^{\alpha}t_2)
=(\sum_{i=1,2}p_it_i^{\frac{1}{\alpha}})^{\alpha}.
\end{eqnarray}

Combining (\ref{delta}), (\ref{max}) and (\ref{min}), we have
\begin{eqnarray*}
\Delta_{\sigma\in \mathcal{I}}
f^{\frac{1}{\alpha}}_{\alpha,z}(\rho,\sigma)=&p_1\Delta_{\sigma_1\in \mathcal{I}}
f^{\frac{1}{\alpha}}_{\alpha,z}(\rho_1,\sigma_1)\\
&+p_2\Delta_{\sigma_2\in \mathcal{I}}
f^{\frac{1}{\alpha}}_{\alpha,z}(\rho_2,\sigma_2).
\end{eqnarray*}
Thus, $C_{\alpha,z}$ satisfies additivity of coherence for
block-diagonal states:
$C_{\alpha,z}(p_1\rho_1\oplus p_1\rho_1)=p_1C_{\alpha,z}(\rho_1)+p_2C_{\alpha,z}(\rho_2).$ $\Box$

$C_{\alpha,z}(\rho)$ actually defines a family of coherence measures which includes several typical coherence measures.

$\bullet$ The coherence $C_{\alpha,z}(\rho)$ with $\alpha=\frac{1}{2}, z=1$, i.e, $C_{\frac{1}{2},1}(\rho)$
is  the coherence $C(\rho)$ of (\ref{C_1}) in Ref. \cite{PRA95.042337}.

$\bullet$
 $\alpha\in(0,1)$ and $z=1$ the coherence $C_{\alpha,1}(\rho)$ is  the coherence $C^{\alpha}_{a}(\rho)$
  in Ref. \cite{98}, where the difference of a constant factor
$\frac{1}{1-\alpha}$ in defining
the coherence has already been taken into
account.

$\bullet$
$\alpha\in(0,1)\cup(1,2]$ and $z=1,$
the coherence $C_{\alpha,1}(\rho)$ is  the coherence $C(\rho)$  in Ref. \cite{R}.

$\bullet$
 $\alpha\in[\frac{1}{2},1)$ and $z=\alpha$; $\alpha>1$ and $z=\alpha,$
the coherence $C_{\alpha,z}(\rho)$ is  the coherence $C_{s,\alpha}(\rho)$  in Ref. \cite{JX}.

In particular, from the relation between the $\alpha$ affinity of coherence \cite{98} and  $C_{\alpha,z}$, we have that $\frac{1}{2}C_{\frac{1}{2},1}(\rho)$
is just the error probability  to discriminate $\{|\varphi\rangle_i,\eta_i\}_{i=1}^{d}$
with von Neumann measurement,  where
$|\varphi\rangle_i=\eta_i^{-\frac{1}{2}}\sqrt{\rho}|i\rangle,$
$\eta_i=\rho_{ii}$ and $d=\sqrt{\rho}$. Furthermore,
if $\rho$ is an incoherent state, the coherence $C_{\frac{1}{2},1}(\rho)=0$, which means that
a set of linearly independent pure states can be perfectly discriminated
by the least square measurement.

\section{The properties of $C_{\alpha,z}(\rho)$}

From Theorem \ref{th1}, $C_{a,1}(\rho)$ is  a well-defined measure of coherence for $\alpha\in(0,1)\cup(1,2]$,
\begin{equation*}
C_{\alpha,1}=\min_{\sigma\in \mathcal{I}}\left[\frac{f^{\frac{1}{\alpha}}
(\rho,\sigma)-1}{\alpha-1}\right],
\end{equation*}
where $f(\rho,\sigma)=tr(\rho^\alpha\sigma^{1-\alpha})$, since for any pair of square matrices $A$ and $B$, the eigenvalues of $AB$ and $BA$ are the same.
For any incoherent state $\sigma=\sum_{k=1}^{d}\delta_{kk}|k\rangle\langle k|$, we have
\begin{eqnarray*}
tr(\sigma^{1-\alpha}\rho^{\alpha})&=&
\sum_{k=1}^{d}\delta^{1-\alpha}_{kk}\langle k|\rho^{\alpha}|k\rangle\\
&=&Q\sum_{k=1}^{d}\frac{\langle k|\rho^{\alpha}|k\rangle}{Q}\delta^{1-\alpha}_{kk},
\end{eqnarray*}
where $Q=\left(\sum_{k=1}^{d}\langle k|\rho^{\alpha}|k\rangle^{\frac{1}{\alpha}}\right)^{\alpha}$.
Denote
\begin{equation*}
\varepsilon(\alpha)=\left\{\begin{array}{llrr}
-1,&0<\alpha<1,\\
1,&1<\alpha.
\end{array}\right.
\end{equation*}

According to the H$\ddot{o}$lder inequality and the
converse H$\ddot{o}$lder inequality, we have
\begin{eqnarray}\label{N}
&&\varepsilon(\alpha)\sum_{k=1}^{d}\frac{\langle k|\rho^{\alpha}|k\rangle}{Q}\delta^{1-\alpha}_{kk}\\\nonumber
&\geq&\varepsilon(\alpha)\left(\sum_{k=1}^{d}\delta_{kk}\right)\left[\sum_{k=1}^{d}\left(\frac{\langle k|\rho^{\alpha}|k\rangle}{Q}\right)^{\frac{1}{\alpha}}\right]^\alpha\\\nonumber
&=&\varepsilon(\alpha),\nonumber
\end{eqnarray}
where the equality is attained when
$\delta^{1-\alpha}_{kk}=\frac{\langle k|\rho^{\alpha}|k\rangle}{Q}$.
Then one finds the following  conclusion.

\begin{corollary}
For $\alpha\in(0,1)\cup(1,2],$
$$C_{\alpha,1}(\rho)=\frac{\sum_{k=1}^{d}\langle k|\rho^{\alpha}|k\rangle^{\frac{1}{\alpha}}-1}{\alpha-1}.$$
And the maximal coherence can be achieved by the  maximally coherent states.
\end{corollary}

That the maximal coherence can be achieved by the  maximally coherent states for $C_{\alpha,1}(\rho)$, with $\alpha\in(0,1)\cup(1,2]$,
can been seen in the following. Based on the eigen-decomposition
of a $d$-dimensional state $\rho=\sum_{j=1}^{d}\lambda_j|\varphi\rangle_j\langle\varphi|$,
with $\lambda_j$
and $|\varphi\rangle_j$ representing the eigenvalue and eigenvectors,
we have:
\begin{eqnarray}\nonumber
\varepsilon(\alpha)\sum_{k=1}^{d}\langle k|\rho^{\alpha}|k\rangle^{\frac{1}{\alpha}}
&=&\varepsilon(\alpha)\sum_{k=1}^{d}\left(\sum_{j=1}^{d}\lambda^{\alpha}_j
|\langle\varphi_j|k\rangle|^{2}\right)^\frac{1}{\alpha}\\\nonumber
&\leq&\varepsilon(\alpha)d^{\frac{\alpha-1}{\alpha}}\left[\sum_{k,j=1}^{d}
\lambda^{\alpha}_j|\langle\varphi_j|k\rangle|^{2}\right]
^\frac{1}{\alpha}\\\nonumber
&=&\varepsilon(\alpha)d^{\frac{\alpha-1}{\alpha}}\left[\sum_{j=1}^{d}
\lambda^{\alpha}_j\right]
^\frac{1}{\alpha},
\end{eqnarray}
where the first inequality is due to
\begin{equation*}
\sum_{k=1}^{n}\lambda_kx_k^{p}\left\{\begin{array}{llrr}
\leq\left(\sum_{k=1}^{n}\lambda_k\right)^{1-p}
\left(\sum_{k=1}^{n}\lambda_kx_k\right)^{p},&0<p\leq1,\\
\geq\left(\sum_{k=1}^{n}\lambda_k\right)^{1-p}
\left(\sum_{k=1}^{n}\lambda_kx_k\right)^{p},&p>1,
\end{array}\right.
\end{equation*}
with $x_k=\sum_{j=1}^{d}\lambda^{\alpha}_j
|\langle\varphi_j|k\rangle|^{2}\geq0,\lambda_k=1$ $(k=1,2,...,n)$ and $p=\frac{1}{\alpha}$.
Then one can easily find that  the upper bound of the coherence  can be
attained by
the maximally coherent states $\rho_d=|\varphi\rangle\langle\varphi|$
with $|\varphi\rangle=\frac{1}{\sqrt{d}}\sum_je^{i\phi_j}|j\rangle$,
$C_{\alpha,1}(\rho_d)=\frac{d^{\frac{\alpha-1}{\alpha}}-1}{\alpha-1}.$
$\Box$

\begin{theorem}
For $\alpha\in(0,1)$, $\beta\in(1,2]$, $\gamma>1$,
$\max\{\alpha,1-\alpha\}\leq z_1\leq1$,
$z_2\geq1$, we have
\begin{equation}\label{21}
C_{\alpha,z_1}(\rho)\leq C_{\alpha,1}(\rho)\leq C_{\alpha,z_2}(\rho);
\end{equation}
\begin{equation}\label{22}
C_{\beta,\beta}(\rho)\leq C_{\beta,1}(\rho)\leq C_{\beta,\frac{\beta}{2}}(\rho);
\end{equation}
And
\begin{equation}\label{23}
C_{\gamma,\gamma}(\rho)\leq
\frac{\sum_{k=1}^{d}\langle k|\rho^{\gamma}|k\rangle^{\frac{1}{\gamma}}-1}{\gamma-1}.
\end{equation}
\end{theorem}

{\sf [Proof]}
Set \begin{equation*}
\varepsilon(z_i)=\left\{\begin{array}{llrr}
-1,&0\leq z_i\leq1,\\
1,&z_i>1,
\end{array}\right.
\end{equation*}
where $i=1, 2$.
According to the Araki-Lieb-Thirring inequality,
for matrixes $A, B\geq0$, $q\geq0$ and for $0\leq r\leq1$, the following inequality holds \cite{AK},
 \begin{equation}\label{z1}
 tr(A^{r}B^{r}A^{r})^{q}\leq tr(ABA)^{rq}.
 \end{equation}
While for  $r\geq1$, the inequality is reversed \cite{AK},
\begin{equation}\label{z2}
 tr(A^{r}B^{r}A^{r})^{q}\geq tr(ABA)^{rq}.
 \end{equation}
From (\ref{z1}) and (\ref{z2}), we have
\begin{eqnarray*}
\varepsilon(z_i)f_{\alpha,z_i}(\rho,\sigma)
&=&
\varepsilon(z_i)tr(\sigma^{\frac{1-\alpha}{2z_i}}
\rho^{\frac{\alpha}{z_i}}\sigma^{\frac{1-\alpha}{2z_i}})^{z_i}\\
&\leq& \varepsilon(z_i)tr(\sigma^{\frac{1-\alpha}{2}}
\rho^{\alpha}\sigma^{\frac{1-\alpha}{2}})\\
&=&\varepsilon(z_i) tr(\rho^{\alpha}\sigma^{1-\alpha})\\
&=&\varepsilon(z_i)f_{\alpha,1}(\rho,\sigma).
\end{eqnarray*}
Combining (\ref{Ca}) and $\alpha\in(0,1)$, we have $C_{\alpha,z_1}(\rho)\leq C_{\alpha,1}(\rho)\leq C_{\alpha,z_2}(\rho).$
(\ref{22}) can be obtained in a similar way.

Since $\gamma>1$, we have $f_{\gamma,\gamma}(\rho,\sigma)\leq tr(\rho^{\gamma}\sigma^{1-\gamma})$.
Similar to the proof of (\ref{N}),
$\min_{\sigma\in I}tr^{\frac{1}{r}}
(\rho^{\gamma}\sigma^{1-\gamma})
=\sum_{k=1}^{d}\langle k|\rho^{\gamma}|k\rangle^{\frac{1}{\gamma}},$
we obtain (\ref{23}). $\Box$

{\it Example 1:}
Let us consider a single-qubit pure state,
$$
\rho=\frac{1}{2}(I_2+\sum_ic_i\sigma_i),
$$
where $\sum_ic_i^2=1$, $I_2$ is the $2\times 2$ identity matrix and
$\sigma_i$ $(i=1,2,3)$ are Pauli matrices.
By Ref. \cite{98}, one has
\begin{equation*}\label{b1}
\max_{\sigma\in \mathcal{I}}tr^2(\sqrt{\sqrt{\sigma}\rho\sqrt{\sigma}})=\frac{1}{2}(1+|c_3|),
\end{equation*}
and
\begin{equation*}\label{b2}
\max_{\sigma\in \mathcal{I}}tr^2(\sqrt{\rho}\sqrt{\sigma})
=\frac{1}{2}(1+c^2_3).
\end{equation*}

For the single-qubit
pure state $\rho$, one has
\begin{eqnarray}
 \rho^{\frac{1}{4}}=\rho=\begin{pmatrix}
\frac{1+c_3}{2}& \frac{c_1-ic_2}{2} \\
\frac{c_1+ic_2}{2}&\frac{1-c_3}{2} \\
  \end{pmatrix}.\quad
\end{eqnarray}
Since $tr(\sigma^{\frac{1}{8}}\rho^{\frac{1}{4}}\sigma^{\frac{1}{8}})^2
=tr(\sigma^{\frac{1}{4}}\rho^{\frac{1}{4}})^2,$
we now compute $\max_{\sigma\in \mathcal{I}}\left[tr(\sigma^{\frac{1}{4}}\rho^{\frac{1}{4}})^2\right]^2.$
Suppose that $\sigma=\sum_ip_i|i\rangle\langle i|$ with $p_1+p_2=1$ and $0\leq p_1, p_2\leq1$.
We have
\begin{eqnarray*}
\sqrt{tr(\sigma^{\frac{1}{4}}\rho^{\frac{1}{4}})^2}
&=&\frac{1+c_3}{2}p_1^{\frac{1}{4}}+\frac{1-c_3}{2}p_2^{\frac{1}{4}}\\
&\leq& \left[\left(\frac{1+c_3}{2}\right)^{\frac{4}{3}}
+\left(\frac{1-c_3}{2}\right)^{\frac{4}{3}}\right]^{\frac{3}{4}},
\end{eqnarray*}
by using the H$\ddot{o}$lder inequality and that the equality
holds if and only $p_1=c(\frac{1+c_3}{2})^{\frac{4}{3}}$
and $p_2=c(\frac{1-c_3}{2})^{\frac{4}{3}}$ with $c=\left[(\frac{1-c_3}{2})^{\frac{4}{3}}
+(\frac{1+c_3}{2})^{\frac{4}{3}}\right]^{-1}.$
Therefore we have
\begin{equation*}\label{b3}
\max_{\sigma\in \mathcal{I}}\left[tr(\sigma^{\frac{1}{4}}\rho^{\frac{1}{4}})^2\right]^2
=\left[\left(\frac{1+c_3}{2}\right)^{\frac{4}{3}}
+\left(\frac{1-c_3}{2}\right)^{\frac{4}{3}}\right]^3.
\end{equation*}

Due to (\ref{Ca}), we obtain
\begin{equation*}
C_{\frac{1}{2},z}(\rho)=2\left[1-\max_{\sigma\in \mathcal{I}}
tr^2(\sigma^{\frac{1}{4z}}\rho^{\frac{1}{2z}}
\sigma^{\frac{1}{4z}})^z\right],
\end{equation*}
then
we have
\begin{equation*}
C_{\frac{1}{2},\frac{1}{2}}(\rho)=1-|c_3|,
\end{equation*}
\begin{equation*}
C_{\frac{1}{2},1}(\rho)=1-c_3^2
\end{equation*}
and \begin{equation*}
C_{\frac{1}{2},2}(\rho)=2-2\left[\left(\frac{1+c_3}{2}\right)^{\frac{4}{3}}
+\left(\frac{1-c_3}{2}\right)^{\frac{4}{3}}\right]^3.
\end{equation*}
It is obvious that
$C_{\frac{1}{2},\frac{1}{2}}(\rho)\leq C_{\frac{1}{2},1}(\rho)\leq
C_{\frac{1}{2},2}(\rho)$, see Fig. 1.
\begin{figure}
  \centering
  \includegraphics[width=7cm]{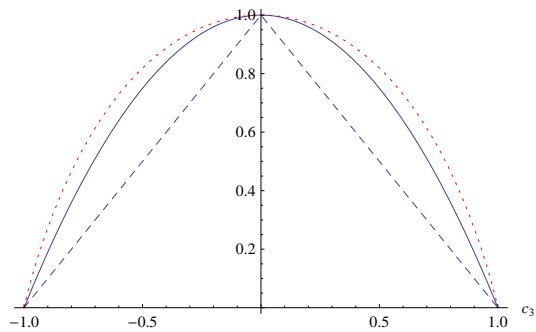}\\
\caption{The red dotted line is the vale of $C_{\frac{1}{2},2}(\rho)$;
The blue solid line is the vale of $C_{\frac{1}{2},1}(\rho)$; The dashed line is the vale of $C_{\frac{1}{2},\frac{1}{2}}(\rho).$
}
\label{fig1}
\end{figure}

\section{conclusion}
In summary, we have proposed four classes of coherence $C_{\alpha,z}(\rho)$
measures based on the generalized $\alpha-z-$relative R$\acute{e}$nyi entropy.
It has been proven
that these coherence measures satisfy all the required
criteria for a satisfactory coherence measure.
Moreover, we have obtained the analytical formulas for special quantifiers with $z=1$
and also studied relations among  the four classes of coherence $C_{\alpha,z}(\rho)$.

\bigskip
\noindent{\bf Acknowledgments}\, \, This work is supported by NSFC under numbers 11675113, 11605083, and Beijing Municipal Commission of Education (KZ201810028042).

\end{document}